\documentclass[showpacs,preprintnumbers,amsmath,amssymb,nofootinbib]{revtex4}
\usepackage{graphicx}
\usepackage{epsfig}
\usepackage{bm}
\usepackage{amsfonts}
\newcommand{\Oo}{{\Delta_2}}

\newcommand{\Po}{{\cal P}}

\newcommand{\eq}{\begin{eqnarray}}
\newcommand{\eqx}{\end{eqnarray}}
\newcommand{\ba}{\begin{equation}}
\newcommand{\ea}{\end{equation}}
\newcommand{\f}[2]{\frac{#1}{#2}}

\newcommand{\cor}[1]{\left\langle{\ #1\ }\right\rangle}

\newcommand{\bit}{\begin{itemize}} 
\newcommand{\eit}{\end{itemize}} 
\newcommand{\ii}{\item}

\def\la{\label}

\def\bi{\bibitem}

\def\va{\varphi}

\begin{document}

\title{On the positivity of Fourier transforms}

\author{Bertrand G. Giraud }
\author{ Robi Peschanski }
\affiliation{Institut de Physique Th\'eorique,\\
CEA, IPhT, F-91191 Gif-sur-Yvette, France\\
CNRS, URA 2306 }
\email{bertrand.giraud@cea.fr; robi.peschanski@cea.fr}

\today

\begin{abstract}
Characterizing in a constructive way the set of real functions whose Fourier
transforms are positive appears to be yet an open problem. Some sufficient
conditions are known but they are far from being exhaustive. We propose two 
constructive sets of  necessary conditions for positivity of the Fourier
 transforms and test their ability of constraining the positivity domain. 
One uses analytic continuation and Jensen inequalities and the 
other deals with 
Toeplitz determinants and the Bochner theorem. Applications are discussed,
 including the extension to the two-dimensional Fourier-Bessel transform and the
 problem of positive reciprocity, i.e. positive functions with positive transforms.
\end{abstract}
\maketitle

\section{Introduction}
\la{introd0}

In a previous paper \cite{gipe}, the following questions were asked:
\bit
\ii
What are the constraints for an even, real function $\psi(r)$ ensuring that its
Fourier transform satisfies,
\ba
\varphi(s) \equiv \sqrt{\frac{2}{{\pi}}}\, \int_{0}^{+\infty}\!\! dr\ 
\cos(s r)\ \psi(r)\ \ge\ 0\, ,
\la{phi}
\ea
\noindent that is, $\varphi$ be real and positive? 
\ii
What are the constraints on even Fourier partners such that $\varphi$ and 
\ba
\psi(r) \equiv \sqrt{\frac{2}{{\pi}}}\, \int_{0}^{+\infty}\!\! ds\ \cos(s r)\ 
\varphi(s) \ \ge\ 0
\, ,
\la{psi}
\ea
be {\it both} positive?
\eit

The motivation for such questions came from various areas of mathematics,
notably \cite{levy}, and from physics, for which a good example is the
Fourier-Bessel relation \cite{kovchegov} between two physically observable
quantities which are necessarily positive (for recent references, see $e.g.$
\cite{math} for maths, and \cite{phys} for various physics domains). In the paper \cite{gipe} an approach based on a 
Gaussian times a combination of a few, low order Hermite polynomials, allowing
to consider eigenfunctions of the Fourier transforms (\ref{phi},\ref{psi}),
has shown that the structure of the positivity space in the manifold of the
components may be quite involved. In the present work we aim at finding
results on a more general basis.

Despite the mathematical interest of the problem, one might argue on practical
grounds that the power of modern computers legitimates a brute force
approach, of the kind, given $\psi$ for instance, numerically calculate
$\varphi$ and ``just look at it''. We strongly feel, however, that, despite
the existence of subtle methods for the integration of oscillatory integrands,
finer analytical methods are in demand. In particular, direct criteria for
$\psi,$ for instance, without any numerical estimates for $\varphi,$ can
provide more elegant and more rigorous approaches.

The main optimal issue for finding out the set of functions solutions would be
an overall $constructive$, sufficient and necessary condition satisfying the
above-mentioned positivity. It seems that one is still far from that goal in
the literature, since it appears to be a rather involved mathematical problem
\cite{eynard}. Indeed some general theorems exist on the positivity of Fourier
transforms \cite{general}, if only the well-known Bochner theorem
\cite{gelfand} on the ``positive-definiteness'' of the function\footnote
{``Positive-definiteness'' of a function $f$ means that for any set
of positions $\{r_i, i=1,...,n\}$ the $n \times n$ matrix with elements
$f(r_i\!-\!r_j)$ is positive definite, i.e. $\sum_{i,j}u_if(r_i\!-\!r_j)u_j\ge 0$ for all test vector $\vec u$.
}
which we will usefully recall later on. They possess a fundamental interest, and will help our studies. However, up to
our knowledge, they still gave only limited practical tools allowing to get
information on the functional set of solutions. Some general useful
 {\it sufficient} constructive conditions are known, such as convexity 
\cite{lafforgue} for the function in the integrand ($i.e.$ $\psi$ in 
\eqref{psi}), but it can be easily shown on specific examples \cite{gipe}
 that it gives only a limited part of the space of solutions. Hence $necessary$
constructive conditions, namely a set of practical constraints on 
functions whose Fourier transforms are positive, would be helpful.

In the present work, we derive two separate sets of such constructive and 
 necessary conditions obeyed by a function, say $\psi(r),$ if its Fourier transform
$\varphi(s)$ is positive. One method (called in short ``matrix method'') is based on the positivity of a hierarchy of
determinants of Toeplitz matrices\footnote{Toeplitz matrices are defined, see
 $e.g.$ \cite{Toeplitz} by  coefficients $c_{ij}\equiv c_{i-j}.$}. The second 
method 
(called ``analytic method'') is based on an analytic continuation onto the 
imaginary axis associated with
 Jensen inequalities \cite{jensen}. They are
 constructive, meaning that they will be
expressed directly (and in a sense ``simply'') in terms of $\psi(r).$ They will 
be formulated as a set of necessary inequalities, for all values of the variable 
$r$ most often.
This we find much constraining, and thus, hopefully, making a step towards a
determination of the set of solutions of the positivity condition \eqref{phi}.
If starting with positive functions $\psi$, the same constraints give useful
indications on the solution of the condition \eqref{psi}, which is our final
goal.

Let us illustrate one typical constraint from the matrix method. Using a generalization of the well-known consequence of Bochner's theorem, 
namely 
\ba
\psi(0) \ge \psi(r)\ ,\quad 
\forall r\ 
\la{2result}
\ea  
coming from the application of positive definiteness to the 2-matrix 
\ba
 \begin{pmatrix}
\psi(0)& \psi(r)\\ \psi(r)& \psi(0) 
\end{pmatrix}\ ,
\la{2matrix}
\ea
 one obtains
\ba
\psi(2r)\ \ge\ 2\ \f{\psi(r)^2}{\psi(0)}-{\psi(0)}\ ,\quad \forall r\ ,
\la{3result}
\ea 
 with one more  dimension of the matrix.
Indeed, considering a
 three-by-three  ``equidistance'' symmetric Toeplitz matrix, 
\ba 
\begin{pmatrix}
\psi(0)& \psi(r)& \psi(2r) \\
\psi(r)& \psi(0)& \psi(r) \\
\psi(2r)& \psi(r)& \psi(0)\\ 
\end{pmatrix}\ ,
\la{3matrix}
\ea
and applying Bochner's theorem, the positive definiteness 
of \eqref{3matrix} gives \eqref{3result}. Noting that \eqref{3matrix} is a 
symmetric 
Toeplitz matrix whose generalization to higher dimensions is straightforward, we
 will take advantage of the hierarchy of  inequalities obtained by increasing
 the   dimensionality. Note also that the positive definiteness of the matrix 
\eqref{3matrix} implies that of (4) using test vectors of the form 
$(a,b,0).$

\begin{figure}[htb] \centering
\mbox{  \epsfysize=37mm
         \epsffile{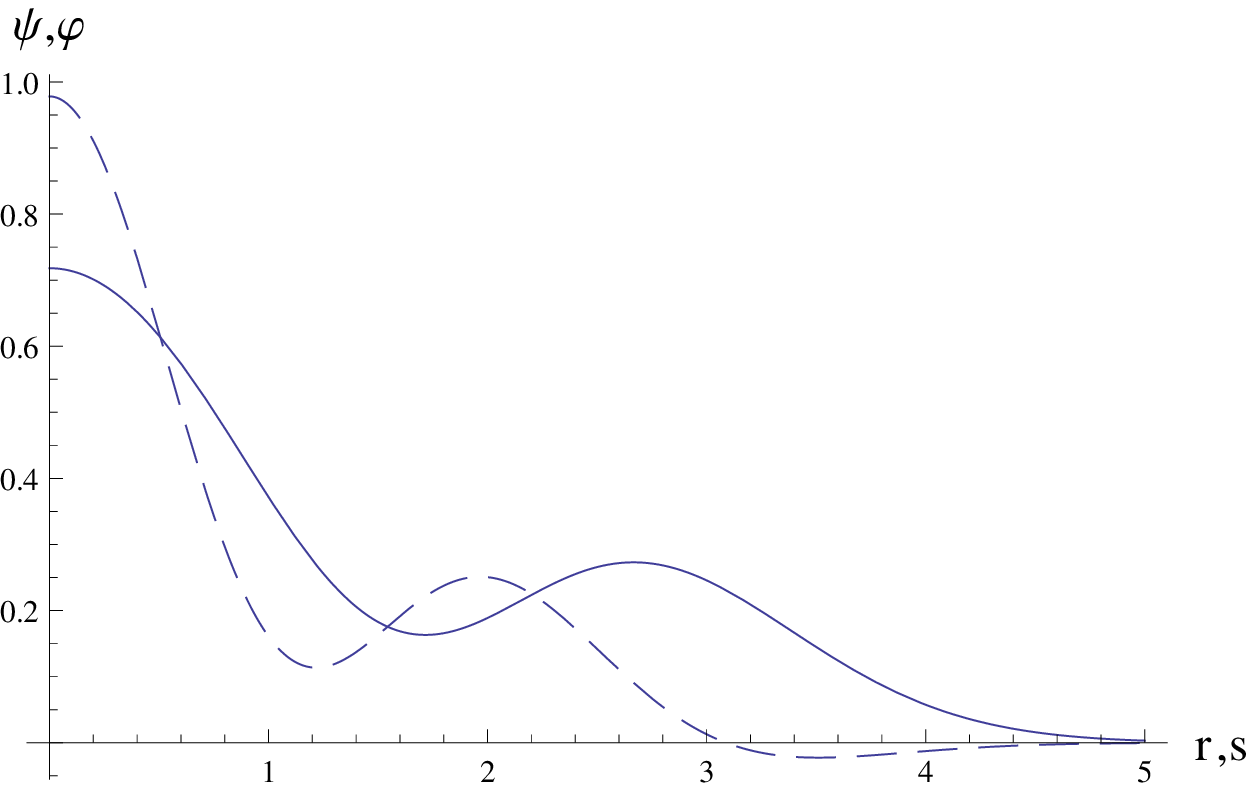}
         \epsfysize=37mm
         \epsffile{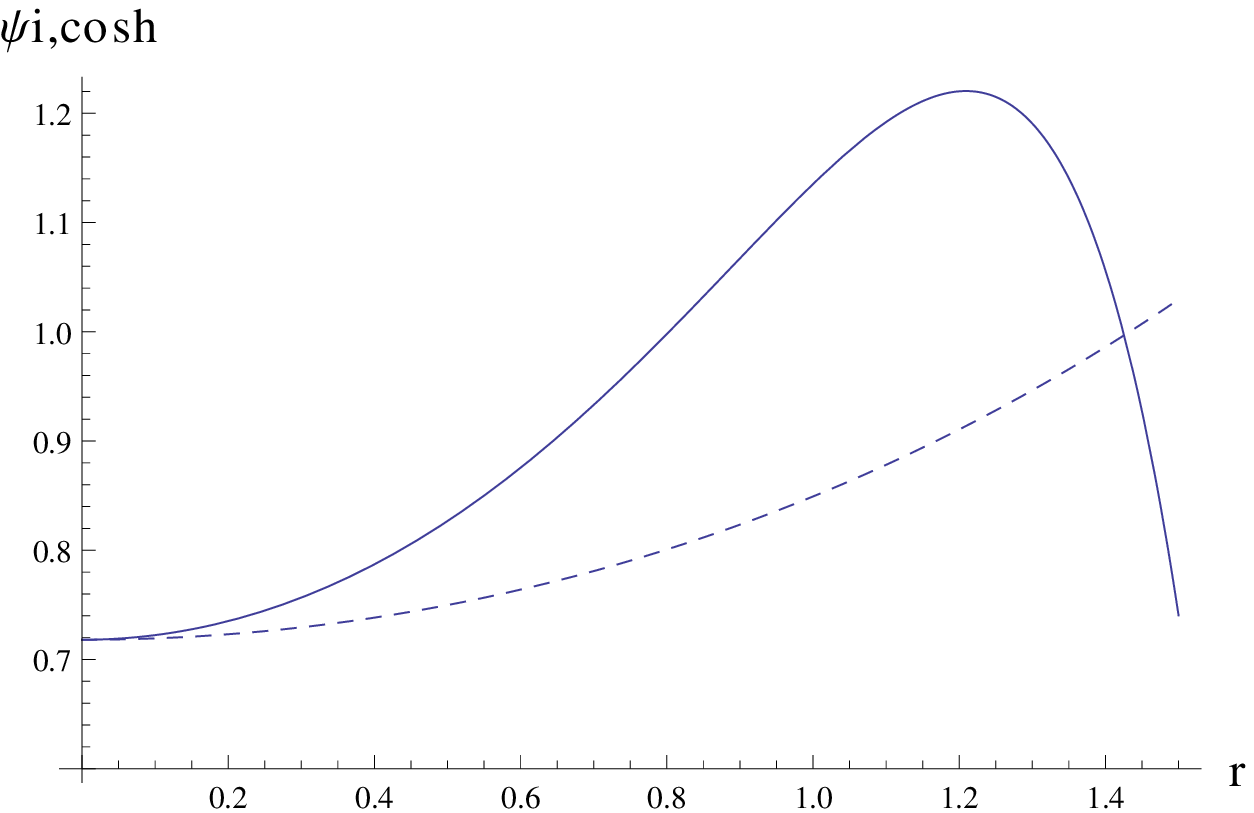}
        \epsfysize=37mm
         \epsffile{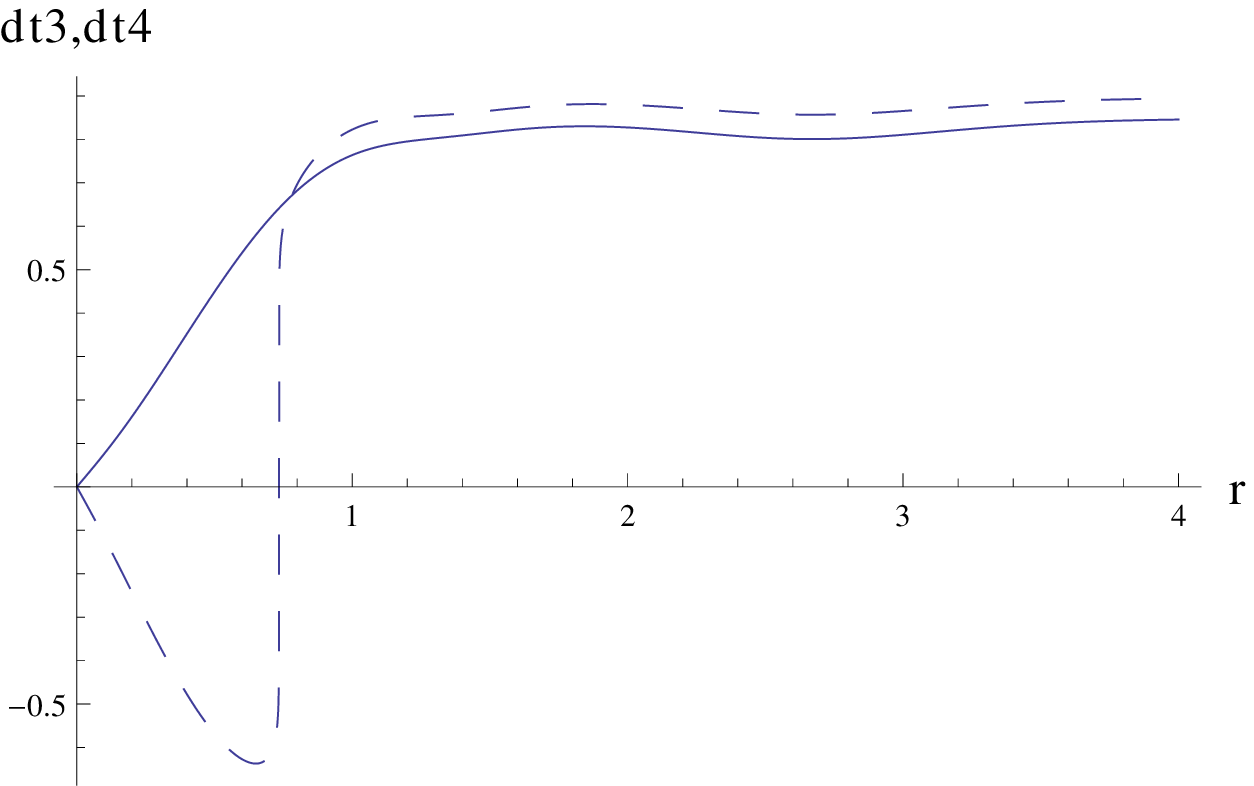}
}
\caption{Left: Full line, positive $\psi(r)$; dashed line, its partly negative
transform $\varphi(s)$. Center: Full line, $\psi i \equiv \psi(ir)$; dashed 
line, its expected $\cosh$ lower bound according to (7) fails, thus detecting the
non-positivity of $\varphi$. Right: Same detection with the matrix method at 
order 4 (dashed line), while order 3 (full line) is insufficient.} 
\end{figure}

To illustrate the ``analyticity method'', one of our results reads,
\ba
\psi({ir})\ \ge\ \psi(0)\ \cosh \left\{ \frac{2\, r}{\pi} \int_0^\infty\!\!
\f {dr'}{r'^2}\ \left(1-\f{\psi(r')}{\psi(0)}\right)\right\}\quad  \forall\ r 
\in [0,\infty[\ ,
\la{constraint}
\ea
where $\psi (ir)$ is the analytic continuation of $\psi(r)$ onto the imaginary
axis. It is a real, convex function, since, by definition \eqref{psi}, $\psi(r)$ is a 
function of $r^2$ and it is assumed here that both integrals, 
$\int_0^{\infty} ds \cosh(s r) \varphi(s),$ and 
$\int_0^{\infty} ds\, s^2 \cosh(s r) \varphi(s),$ are convergent. As will be shown in the following, the coefficient of $r$
in the $\cosh$ seen in formula (3) is the average value of $s$ in Fourier
space, but the formula, obviously written in the $\psi$ space, introduces no
oscillating cosine term. 
Figure 1 illustrates a typical case\footnote{
The function and its Fourier transform read,
$\psi(r)= e^{-\frac{r^2}{2}}
(0.718081 - 0.064879 r^2 - 0.0685793 r^4 + 0.0269736 r^6 + 0.00119983 r^8)$,
$\varphi(s)= e^{-\frac{s^2}{2}}
(0.97805 - 1.24138 s^2 + 0.587989 s^4 - 0.0605688 s^6 + 0.00119983 s^8)$;
the average of $s$ is  $\langle s \rangle=0.836263$.
} where the ``cosh test'' \eqref{constraint}  eliminates a candidate $\psi$,
if only because $\psi(ir)$ is blatantly non convex,
and the same detection is obtained provided we use a Toeplitz matrix of order 4
or more.

The plan of the paper is as follows. The next section Sec. II, is devoted to the
 description of the two sets of 
constraints, namely those coming from the matrix method, and then from the
 analyticity method.  Section III provides a numerical
investigation of the compared efficiency of both methods based on the approach 
initiated in Ref.\cite{gipe}. Section IV  extends
our reasoning and numerical evaluation to the 2-dimensional case of the 
Fourier-Bessel transform. Section V is a summary and conclusion.

\section{Necessary Conditions}
\la{Introd}

\subsection{Simple positivity and maximality tests}

In the following, $\psi(r)$ is an even function, positive,
($\psi(r) \ge 0,\ \forall r$), and we further assume that $\psi$ has as many
derivatives as might be needed. If its Fourier partner $\varphi$ is also
positive, then, {\it necessarily},

\medskip
i) The function $\psi$ is ``positive definite'', hence $\psi$ {\it is 
maximum at the origin}, namely $\psi(0) \ge \psi(r),\ \forall r,$ see \eqref{2result}. This may be seen 
from the Bochner theorem for the pair of points, $x_1=0,\, x_2=r$, and the
test vector, $\vec u=(1,-1)$, leading to
\begin{equation}
\sum_{i,j=1.2} u_i\, \psi(x_i-x_j)\, u_j=2[\psi(0)-\psi(r)]\ge\, 0\ .
\end{equation}

\medskip
ii) For every positive integer $q$ every product, 
$\varphi_q \equiv s^q \varphi$, is positive, hence the Fourier partner,
$\psi_{2q} \equiv (-)^q\, d^{2q}\psi/dr^{2q}$, taking into account the change
of sign in front of the cosine in formula (2) verifies the same maximality 
condition $i.e.$ 
$\psi_{2q}$ {\it has its maximum at the
origin}.

\medskip
iii) Every moment is positive, 
$\mu_q \equiv \int_0^{\infty} ds\, s^q \varphi(s) >0$, hence in particular,
for {\it even} moments and thus {\it from every even derivative of $\psi$, the 
condition} $\psi_{2q}(0)>0$. The somewhat more involved expression for 
{\it odd} moments is derived in Subsection C, see further.

\medskip
iv) {\it Every average value}, such as $\mu_q/\mu_0$ or 
$\mu_q/\mu_{q-1}$, {\it is positive}. We show in the following how,
while even moments, see iii), are easily observed in the $r$-space, odd
average values are also available from the $r$-space.

\medskip
v) The positivity, for any width $b$ in momentum space, of a product,
$e^{-s^2/(2b^2)}\, \varphi$, of $\varphi$ by a Gaussian, induces the
``positive-definiteness'' of the result of the corresponding convolution,
\begin{equation} 
\psi_b(r) \equiv  \f b{\sqrt{2\pi}}\int_0^{\infty} dr'\ \left[\ e^{-b^2(r-r')^2/2}
+e^{-b^2(r+r')^2/2}\ \right]\ \psi(r')\ ,
\label{convolu}
\end{equation}
with, {\it for} $\psi_b$ {\it and its derivatives, all the consequences
already seen under} i)-iv).

\subsection{Matrix method}

\medskip
Elaborating on the positive definiteness of the function $\psi$ following from the Bochner theorem,
let us introduce the hierarchy of symmetric real Toeplitz n-matrices $T_n$
\ba \begin{pmatrix}
\psi(0)& \psi(r)& \psi(2r) & ... & \psi[(n\!-\!1)r]\\
\psi(r)& \psi(0)& ...& ...& \psi[(n\!-\!2)r]\\
...    &     ...& ...&          ...& ...\\
 \psi[(n\!-\!1)r]& \psi[(n\!-\!2)r]& ...& \psi(r)& \psi(0)\\ 
\end{pmatrix}\ .
\la{Nmatrix}
\ea
Application of the Boechner theorem can be shown to lead to the following set 
of 
 inequalities on Toeplitz determinants of positive-definite functions (thus whose Fourier transforms are positive):
\ba
 ||T_n|| > 0,\quad \forall n\ .
\la{determinants}
\ea
It is known that the matrices $T_{n-1}$ and $T_n$ have intertwined eigenvalues,  Hence the smaller eigenvalue of $T_{n-1}$ is always larger than the one of $T_n,$
which is itself positive from positive definitiveness. Hence verifying  \eqref{determinants} for any given $n,$ ensures positivity\footnote{Note also the convergence of  $||T_n||^{1/n} ,\ n\to\infty,$ to a well-defined Fourier series through the Szeg\"o-Ka\^ c theorem, see $e.g.$ \cite{Toeplitz}.} of all determinants $||T_k||$ for $k < n.$ Indeed, the matrix method will make a practical use of \eqref{determinants}
for 
finite $n.$ 
Notice, however, that $||T_n||$ vanishes like $\propto r^{n(n-1)}$ when 
$r \rightarrow 0$, creating numerical accuracy problems for its sign. It may
be more efficient to track the first change of sign of the lowest eigenvalue.

\subsection{Analyticity method}
\medskip
Recall that $\mu_0 \equiv \int_0^{\infty} ds\, \varphi(s)= \sqrt{\pi/2}\,
\psi(0)$. Consider the probability distribution, 
${\cal P}(s) = \sqrt{2/\pi}\, \varphi(s)/\psi(0)$. By analytic continuation of
(2) in $r$-space we obtain,
\begin{equation} 
\psi(i r)=\psi(0) \int_0^{\infty} ds\, \cosh(sr)\, {\cal P}(s) = \psi(0)\, 
\langle\, \cosh(sr)\, \rangle\, ,
\la{imaginary}
\end{equation}
where we have already assumed that $\varphi(s)$ decreases fast enough when 
$s \rightarrow \infty$ to ensure convergence of the integral. Then our key
ingredient is to make use of the well-known Jensen inequality \cite{jensen}, 
$\langle\, f(X)\, \rangle_{\cal P} \ge f(\, \langle X \rangle_{\cal P}\, ),$
valid for any probability distribution ${\cal P}({ X}),$ and any 
{\it convex} function $f$, the hyperbolic cosine being obviously convex.
{\it The following lower bound holds}, 
\begin{equation}
\forall r,\  \psi(ir) \ge \psi(0)\, \cosh(\, \langle s \rangle\, r)\, .
\la{inequality}
\end{equation}
The calculation in $r$-space of $\langle s \rangle = \mu_1/\mu_0$ is 
explained in the next subsection D.

\medskip
An extension of the inequality \eqref{inequality} can be obtained as follows. For notational simplicity, use a short notation $\sigma$ for 
$\langle s \rangle$ and, temporarily, set $\psi(0)=1,$ an inessential
normalization. More inequalities are easily found if one notices that
\eq
\psi(r)+\psi(ir)= \langle\, [\cos(rs)+\cos(irs)]\, \rangle \ge
[\cos(r \sigma )+\cos(i r \sigma )], \\
\psi(r)-\psi(ir)= \langle\, [\cos(rs)-\cos(irs)]\, \rangle \le
[\cos(r \sigma )-\cos(i r \sigma )],
\label{combi4}
\eqx
since the sum, $[\cos x + \cos(ix)]$ and the difference, $[\cos x - \cos(ix)]$,
are obviously a convex and a concave real function of $x$, respectively.

Equivalently, consider now the first complex eighth root $\omega$ of unity,
$\omega^2=i$. It is trivial to find that the combinations,
$[\cos x+\cos(\omega x)+\cos(\omega^2 x)+\cos(\omega^3 x)]$ and 
$[\cos x-\cos(\omega x)+\cos(\omega^2 x)-\cos(\omega^3 x)]$ are real convex
and that
$[\cos x-i \cos(\omega x)-\cos(\omega^2 x)+i \cos(\omega^3 x)]$ and 
$[\cos x+i\cos(\omega x)-\cos(\omega^2 x)-i \cos(\omega^3 x)]$ are real
concave. The following inequalities result,
\eq
\psi(r)+\psi(\omega r)+\psi(ir)+\psi(\omega ir) \ge 
\cos(\sigma r)+\cos(\omega \sigma r)+\cos(i \sigma r)+\cos(\omega i \sigma r),
\\
\psi(r)-\psi(\omega r)+\psi(ir)-\psi(\omega ir) \ge 
\cos(\sigma r)-\cos(\omega \sigma r)+\cos(i \sigma r)-\cos(\omega i \sigma r),
\\
\psi(r)-i \psi(\omega r)-\psi(ir)+i \psi(\omega ir) \le \cos(\sigma r)
-i \cos(\omega \sigma r)-\cos(i \sigma r)+i \cos(\omega i \sigma r),
\\
\psi(r)+i \psi(\omega r)-\psi(ir)-i \psi(\omega ir) \le \cos(\sigma r)
+i \cos(\omega \sigma r)-\cos(i \sigma r)-i \cos(\omega i \sigma r),
\label{combi8}
\eqx
Further roots of unity clearly give further inequalities.

\medskip
The bounds found in (\ref{inequality}-\ref{combi8}) obviously hold, {\it mutatis 
mutandis}, for $\psi_b$ and even derivatives of $\psi$ and $\psi_b$.

\subsection{Odd Moments}

Odd average values, appearing as inputs in inequalities 
(\ref{inequality}-\ref{combi8}), are obviously
also required to be positive, as previously noted as condition iii). It is
important to obtain them as simply as possible in terms of $\psi(r)$. However,
this is less obvious since odd moments of $\varphi$ cannot be trivially
observed from mere derivatives of $\psi$.

Let us derive them using the Laplace transform of $\varphi$,
\ba
\chi(r) \equiv  \int_0^\infty \!\!ds\ e^{-rs}\ \va(s)\ \Rightarrow\ 
\int_0^\infty\!\!  ds\, s\, \va(s) = - \f {d\chi}{dr}(r\!=\!0),
\la{laplace}
\ea
which allows a useful expression of the first moment of $\varphi$.
Indeed, starting with the expression,
\ba
\chi(r)=\sqrt{\f 2\pi}  \int_0^\infty\!\! ds\  e^{-rs}\ \int_0^\infty \!\!dr'
\psi(r')  \cos{sr'} \ ,
\la{laplacebis}
\ea
for $r \ne 0$ one can switch the order of integrations in \eqref{laplacebis}
and write, 
\ba
\chi(r)= \sqrt{\f 2\pi} \int_0^\infty \!\!\psi(r')\, dr'  \int_0^\infty 
\!\!ds\ e^{-rs}\cos{sr'}= \sqrt\f 1{2\pi} \int_0^\infty \!\!dr' 
\left\{\f 1{r+ir'}+\f 1{r-ir'}\right\}\psi(r')\ .
\la{laplaceter}
\ea 
By integration by part we then obtain,
\ba
\chi(r)=  i \sqrt{\f 1{2\pi}} \int_0^\infty \!\!dr'  
\log\left\{\f {r+ir'}{r-ir'}\right\}\f {d\psi(r')}{dr'},
\la{laplaceterbis}
\ea 
since the boundary term,
\ba
-i\sqrt{\f 1{2\pi}} \left[\log\left\{\f {r+ir'}{r-ir'}\right\}\ \psi(r')
\right]_{r'=0}^{r'=\infty}\ ,
\la{first}
\ea
obviously vanishes for all $r > 0.$

Actually, the expression \eqref{laplaceterbis} of gthe function $\chi,$ well defined also at $r\!=\!0$ by continuity,
is the correct analytic continuation of formula \eqref{laplaceter} at 
$r\!=\!0$ and thus the representation of the regular Laplace transform 
\eqref{laplacebis} for all $r\ge 0$.

Finally, one gets the following  formula for the first moment displayed in 
\eqref{laplace},
\ba
 \int_0^\infty\!\!\!\!
 s{ds}\  \va(s) \equiv - \f {d\chi}{dr}(r\!=\!0)\ = \ -\ \sqrt{\f 2\pi}\ 
\int_0^\infty \!\!\f {dr'}{r'}\ \f {d\psi(r')}{dr'}\ .
\la{laplace4}
\ea
It is also possible to make a ``reverse'' integral by part, using an 
$a\ priori$ arbitrary integration constant $C,$ namely,
\ba
\int_0^\infty\!\!\!\!  s{ds}\  \va(s) = 
-\sqrt{\f 2{\pi}} \left\{\f 1 {r'}\ \left[\psi(r')-C\right]
\right\}_{r'=0}^{r'=\infty} - \sqrt{\f 2{\pi}} \int_0^\infty 
\!\!\f{dr'}{r'^2} \left[\psi(r')-C\right] .
\la{laplace5}
\ea 
In fact, the boundary term is nonsingular (and finally zero) only when
choosing the constant $C = \psi(0)$, leading to the final formula,
\ba
\int_0^\infty\!\!\!\!
 s\,{ds}\  \va(s) =   \sqrt{\f 2{\pi}} \int_0^\infty \!\!\f{dr'}{r'^2} 
\left[\psi(0)-\psi(r')\right]\ .
\la{finalpart}
\ea
Inserting the result \eqref{finalpart} into the inequality \eqref{inequality}
gives rise to formula \eqref{constraint}, a self-content functional expression
depending only on the knowledge of $\psi.$ We stress again that the formula
involves no oscillating terms, except those which might be present in $\psi.$

It will be noticed, incidentally, that any numerical result from 
Eq. \eqref{finalpart} that returns a negative value, $\mu_1 <0,$ for this
moment, negates at once the expected positivity of $\varphi$. A detailed
comparison between $\psi(ir)$, $\psi(r)\pm \psi(ir)$, etc and their
respective bounding functions, as detailed above, becomes useless in such
a case of negative moment.

Higher order odd moments, $\mu_{2q+1}$ are obviously obtained upon
substituting $\psi_{2q}$ for $\psi$ in Eq. \eqref{finalpart}. And the
generalization of Eq. \eqref{finalpart} to $\psi_b$ and its moments is
trivial.

In the following Section III we show numerical, illustrative cases, the existence of so-called ``rebels'', where, although 
$\varphi$ contains negativity, several among the known criteria described previously
 are not
violated. Then we test and compare in a systematic way the efficiency of both methods in creating an efficient 
``filter'' for positivity.

\section{Numerical investigations}

\subsection{1-d test functions with two parameters;  maximality and cosh tests efficiency (Subsecs. II A and C)}

We first consider combinations, $\psi=c_0\, \xi_0+c_2\, \xi_2+c_4\, \xi_4,$ 
of the following three harmonic oscillator eigenstates,
\begin{equation}
\xi_0=\pi^{-\frac{1}{4}}\, e^{-r^2/2},\ \ \xi_2=\pi^{-\frac{1}{4}}\, 
e^{-r^2/2}\, (2 r^2-1)/\sqrt{2},\ \ \xi_4=\pi^{-\frac{1}{4}}\, e^{-r^2/2}\,
(4 r^4-12 r^2+3)/(2 \sqrt{6}),
\end{equation}
with $c_0=\cos \alpha ,\ \ c_2=\sin \alpha \cos \beta ,\ \ 
c_4=\sin \alpha \sin \beta ,\ \ 0 \le \alpha < \pi/2$\ \ and\ \ 
$0 \le \beta < 2 \pi$. We explored the
corresponding half sphere via a grid of parallels and meridians, separated
from each other by steps of $\pi/90$ and $\pi/45$ respectively. Only those
functions $\psi$ that are positive are retained for further studies. This 
selects 951 grid points, out of which 520, shown as tiny dots in all parts
of Figures 2-4, correspond to simultaneous $\psi>0$ and $\varphi>0$. The 
remaining 431 points, shown with larger dots in the left part of Fig. 2,
correspond to those cases where $\varphi$ has negative parts. 

{\it Preselection by conditions of maxima at the origin} : 
As stated in Subsec. II A, it is easy to
eliminate 271 out of these 431 cases, via a double maximality condition,
namely that $\psi(0)$ and $\psi_2(0)$ be maxima for $\psi$ and $\psi_2$,
respectively. The 160 remaining ``rebels''  satisfy this double maximality
condition. 

The right part of Fig. 2 shows, besides the 520  ``fully
acceptable cases'' ({\it tiny dots)}, 160  cases where the 
non-positivity of $\varphi$ is $not$ detected by the double maximality 
condition ({\it moderate size dots}) and 271  cases where the same condition is 
efficient for the detection ({\it big dots}).

\begin{figure}[htb] \centering
\mbox{  \epsfysize=60mm
         \epsffile{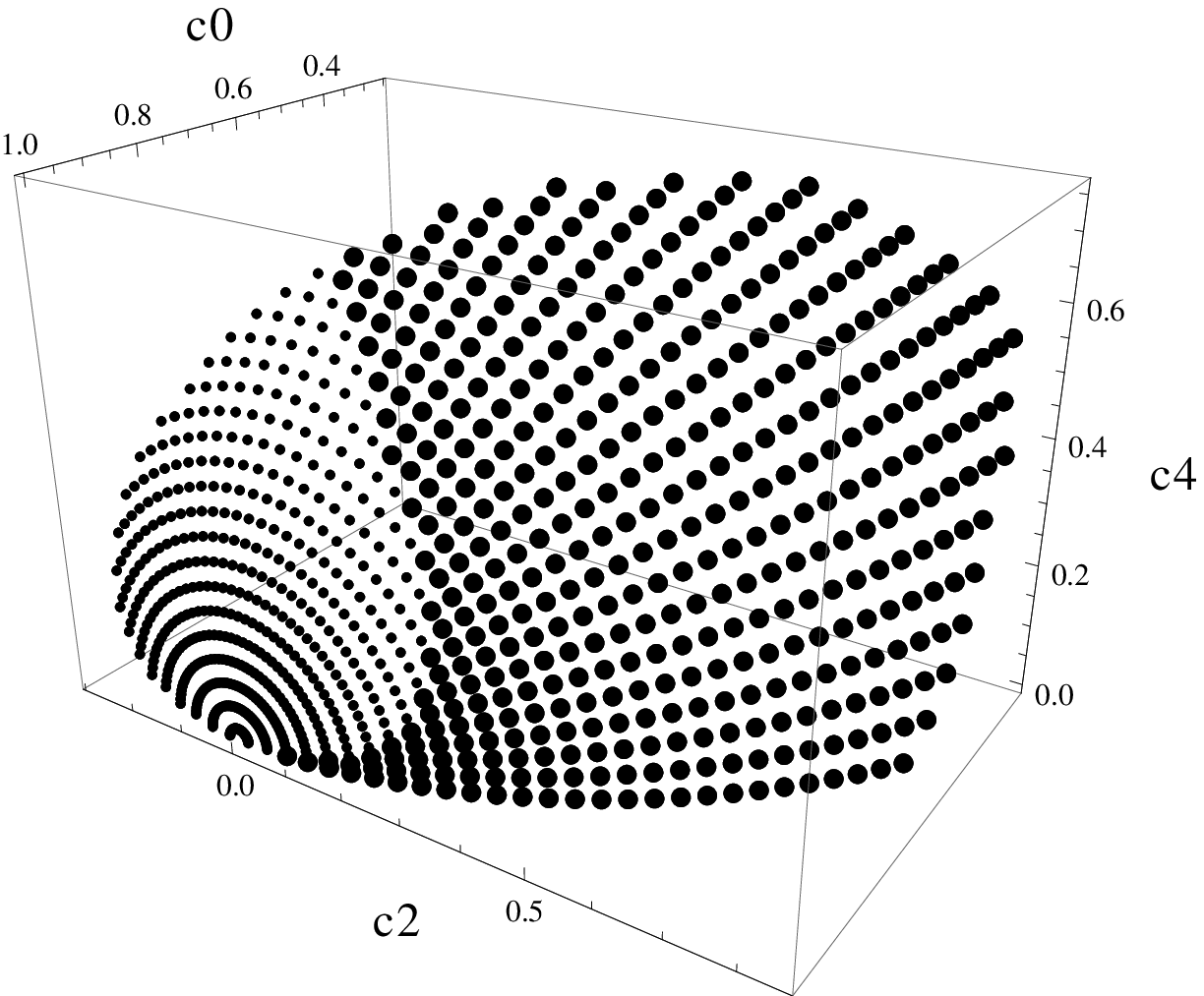}
\hspace{1.0cm}
        \epsfysize=60mm
         \epsffile{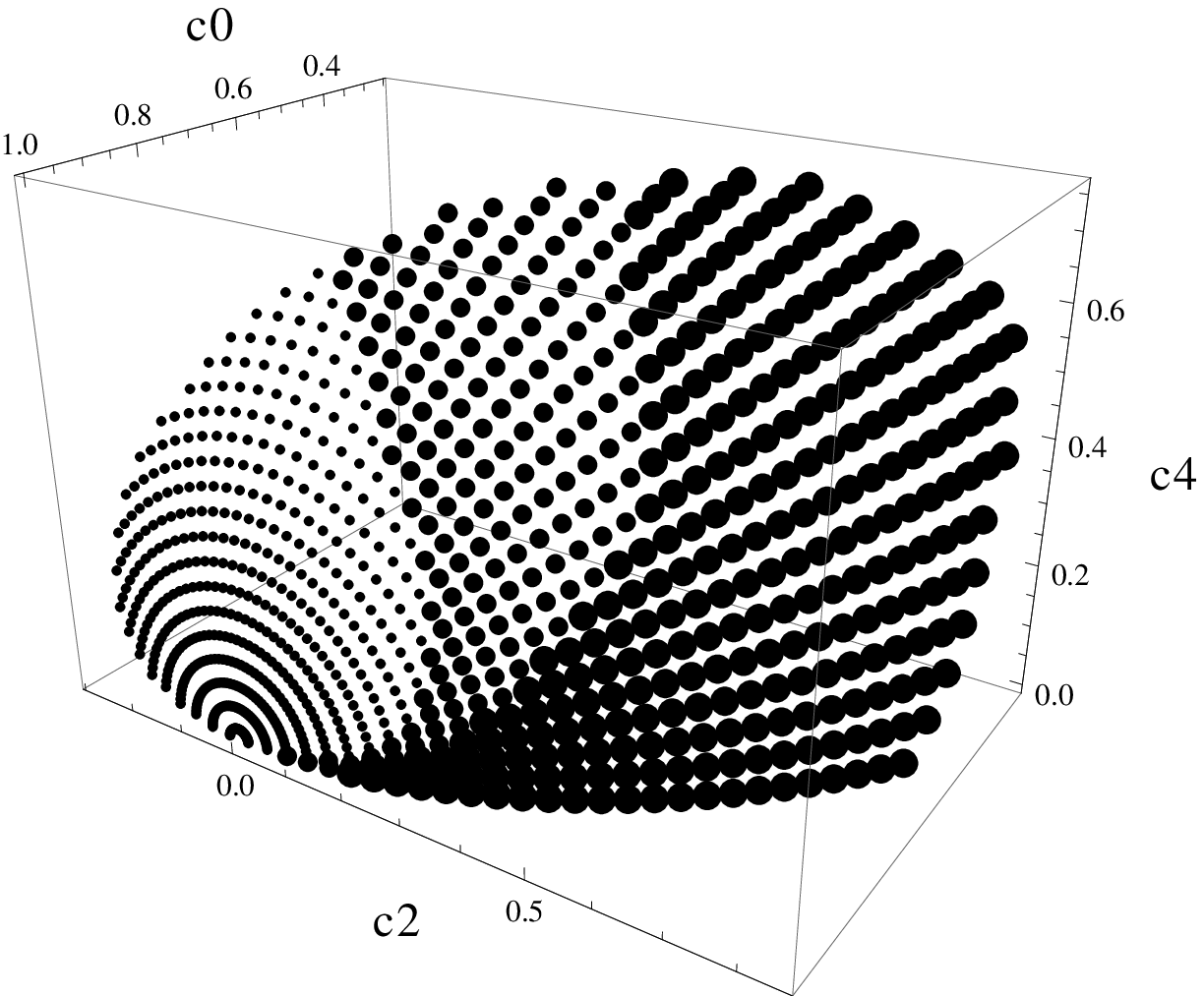} 
}
\caption{Left: Tiny dots, simultaneous $\psi>0$, $\varphi>0$. Bigger dots, 
$\psi>0$, but non-positivity of $\varphi$. Right: Same tiny dots. Big dots,
cases eliminated by maximality conditions for $\psi(0),\, \psi_2(0)$.
Moderate dots, non-positivity undetected by these conditions.}
\end{figure}

\begin{figure}[htb] \centering
\mbox{  \epsfysize=60mm
         \epsffile{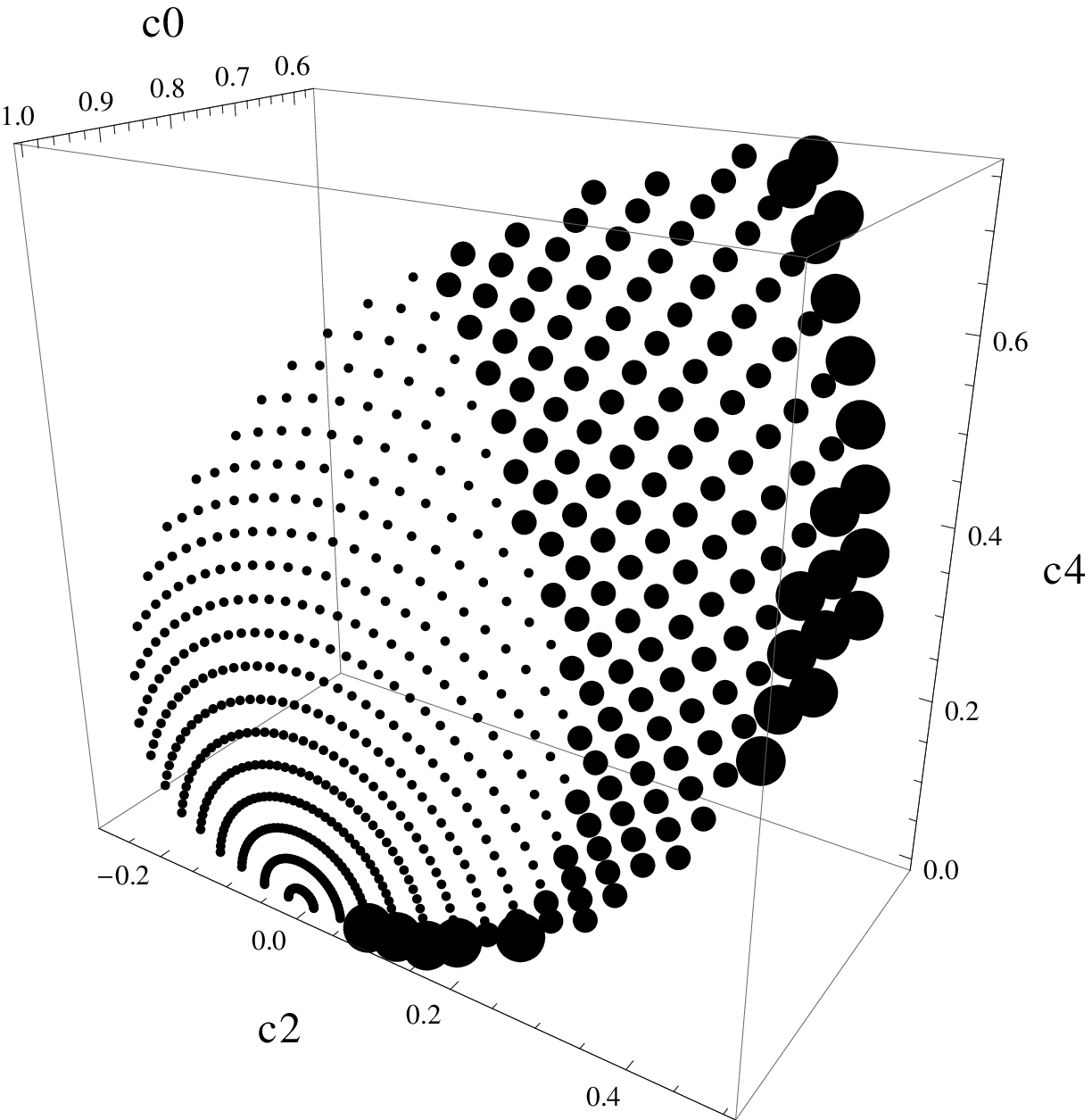}
\hspace{1.0cm}
        \epsfysize=60mm
         \epsffile{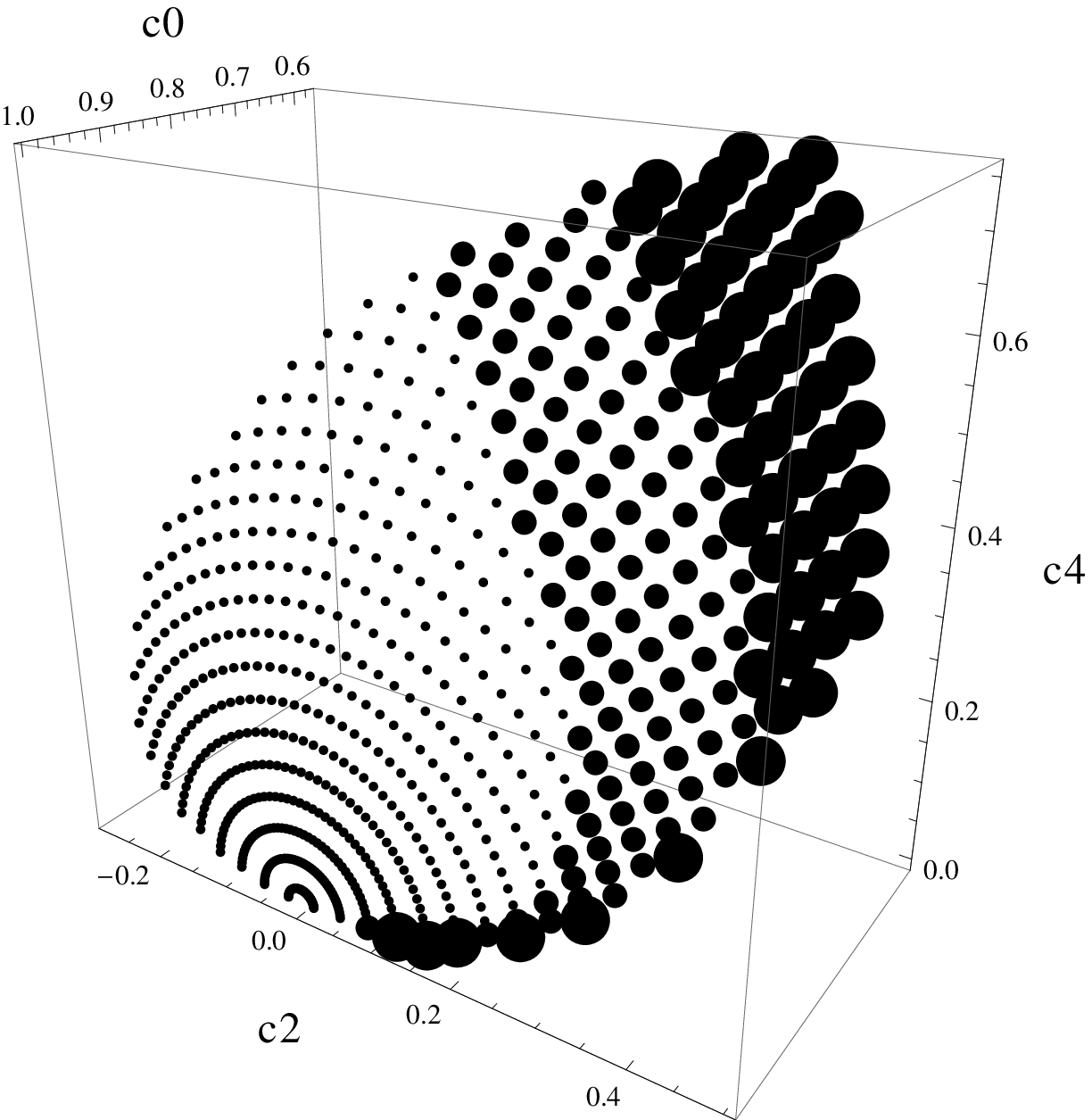}
}
\caption{Tiny dots, simultaneous $\psi>0$, $\varphi>0$. Big dots, $\psi>0$,
non-positivity of $\varphi$ detected with convolution parameter $b=2$, left,
or $b=1,$ right. Moderate dots, $\psi>0$, non-positive $\varphi_b$ undetected.}
\end{figure}

{\it Further screening by a combination of maximality  and cosh tests, including convolution} :
Figure 3 shows, for those 160 functions $\psi$ for which $\varphi$ is 
non-positive although $\psi(0)$ and $\psi_2(0)$ are maxima, the result of an
analysis of their ``convoluted'' $\psi_b$, see Eq. \eqref{convolu}, with 
$b=2$, then b=1, for the left, then right, part of the Figure, respectively. 
The following eight conditions are used, simultaneously : 
a) $\psi_b(0)>0$, $\psi_b(0)$ maximum of $\psi_b$, b) the same two 
conditions for $\psi_{b2} \equiv -d^2 \psi_b/dr^2$, c) both average 
moments for $\psi_b$ and $\psi_{b2}$ must be positive, d) both analytic
continuations $\psi_b(ir)$ and $\psi_{b2}(ir)$ must be larger than
their respective ``$\cosh$'' lower bound. (Note, incidentally, that
preselected functions $\psi$, so that $\psi(0)>0$ and $\psi_2(0)>0$ make
maxima, do not extend necessarily such a positivity and maximality to 
$\psi_b(0)$ and $\psi_{b2}(0)$.) The big dots in Fig. (3) indicate when one,
or several violations, eliminate some among the 160 non-positive $\varphi$'s.
It turns out that the choice, $b=1$, is here more efficient, with 58
functions rejected out of 160, that the choice, $b=2$, with only 23 
detections. An even better result is obtained if $b=1/2$, with 83 detections. 
Clearly, with a more general set of test functions, it will be 
advisable to use a multiplicity of values of $b$ and more criteria to make 
the detection very efficient, the price of further calculations following this method having to be paid.

{\it Moderate efficiency of the ``cosh bound'' criteria (cf.\ref{inequality}-\ref{combi8})} : 
The efficiency of the criterion, 
$\psi(ir) \ge \psi(0)\, \cosh \left(\, \langle s \rangle\, r \right)$, 
even when it is reinforced by its generalizations for $\psi_2$, $\psi_b$ and 
$\psi_{b2}$, turns out to be slightly disappointing for the present set
of 160 ``rebel'' functions. Most detections can already come from the simpler
positivity and maximality criteria for $\psi_b,\, \psi_{b2}$ and so on.
It seems, however, that the criterion is systematically quite useful for
somewhat subtle cases where the negative parts of $\varphi$ are very weak,
but extend to high momenta. To attempt an improvement of the criterion,
define then the ``complement'' function, $\psi_c \equiv \psi-\psi_b$. the
Fourier transform of which is, obviously, $(1-\exp[-s^2/(2b^2)])\, \varphi$.
Distinct average momenta, $\langle s \rangle_b$ and $\langle s \rangle_c$
 correspond to $\psi_b$ and $\psi_c$, respectively. A more precise bound
results,
\begin{equation}
\psi(ir) \ge \psi_b(0)\, \cosh \left[\, \langle s \rangle_b\, r\, \right] +
             \psi_c(0)\, \cosh \left[\, \langle s \rangle_c\, r\, \right]\, .
\end{equation}
It is then trivial to extend this process to more components for the bound.
For instance one could use two values $b$ and $b'$ simultaneously for the 
convolution parameter, hence test functions $\psi_b$ and $\psi_{b'}$, and set
a combination, $\psi_c=\psi-w\, \psi_b-w'\, \psi_{b'}$, with positive
weights $w, w'$ such that $w+w'\le 1,$ to retain the assumed positivity of
the respective Fourier images $\varphi_b,\, \varphi_{b'},\,\varphi_c$\, .
The corresponding bound,
\begin{equation}
\psi(ir)\ge w\,  \psi_b(0)\,    \cosh \left[\, \langle s \rangle_b\, r\, 
\right]  +  w'\, \psi_{b'}(0)\, \cosh \left[\, \langle s \rangle_{b'}\, r\, 
\right]  +       \psi_c(0)\,    \cosh \left[\, \langle s \rangle_c\, r\,
\right]
\, .
\end{equation}
obviously samples the ``probability distribution'' ${\cal P}(s)$ better.

{\it Criteria from convolution and even derivatives} :
Alternately, a multiplication of $\varphi$ by a product of the form, 
$s^{2q} \exp[-s^2/(2b^2)]$, obviously combines derivatives and convolution, to
provide further probes of $\varphi$ in adjustable domains, and better detect
possible defects of positivity.

\begin{figure}[htb] \centering
\mbox{  \epsfysize=60mm
         \epsffile{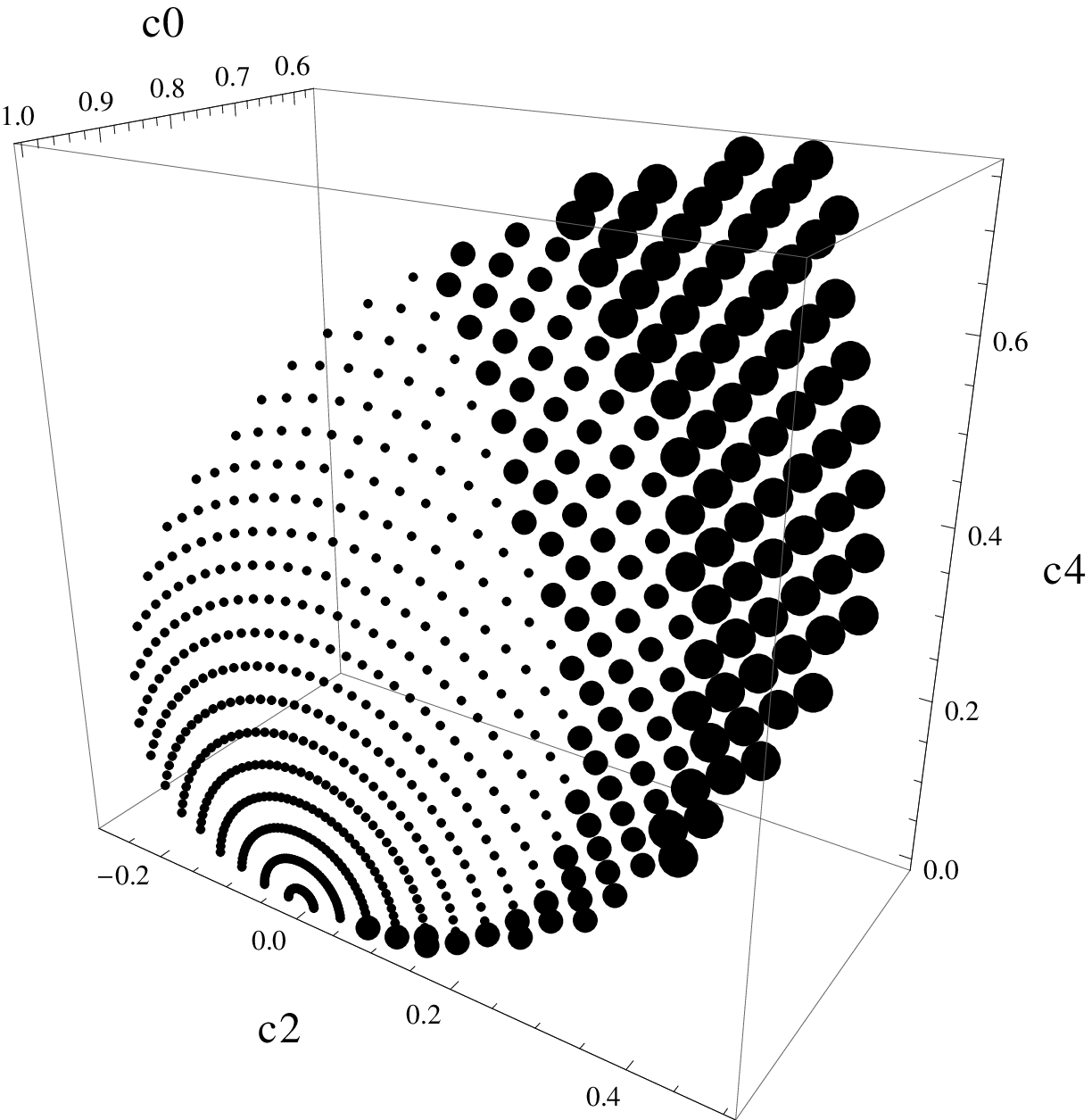}
\hspace{1.0cm}
        \epsfysize=60mm
         \epsffile{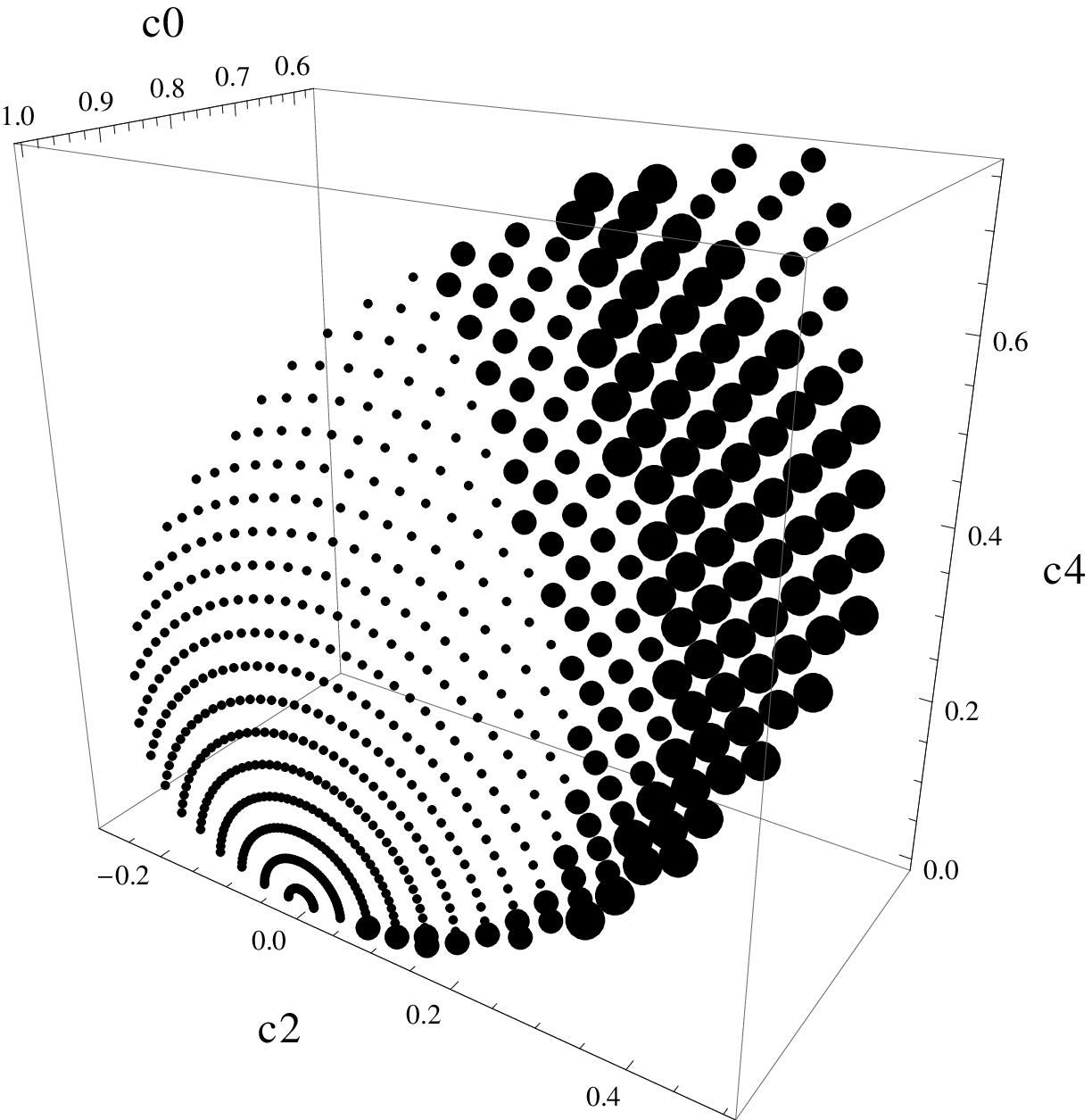} 
}
\caption{Tiny dots, simultaneous $\psi>0$, $\varphi>0$. Big dots, $\psi>0$,
non-positivity of $\varphi$ detected with $b=1/2$ and just the hyperbolic
cosine bound for $\psi_{b2}$, left, or for $\psi_{b4}$, right. Moderate dots,
$\psi>0$, non-positive $\varphi$ undetected by such a bound alone.}
\end{figure}

Recall that 
$\psi_{b2}(r)=-b/\sqrt{2\pi}\ d^2/(dr^2) \int_0^{\infty} dr'[e^{-b^2(r-r')^2/2}
+e^{-b^2(r+r')^2/2}]\psi(r')$, with a similar formula for $\psi_{b4}$.
The left part of Figure 4 shows, with $\, b=1/2$, how the bound, 
$\psi_{b2}(i r) \ge \psi_{b2}(0)\, \cosh \left[\, \langle s \rangle_{b2}\, r\,
\right]$, taken alone, converts 79 among the 160 ``moderate size'' dots into 
``big dots''. Its right part shows the same, with 80 detections, for the
bound, 
$\psi_{b4}(i r) \ge \psi_{b4}(0)\, \cosh \left(\, \langle s \rangle_{b4}\, r\,
\right)$. These two sets of big dots, while much overlapping, show significant
differences. If both criteria are used simultaneously, the detection rate
increases to $94/160 \simeq 59\%$. 

Returning to combining criteria already
used in this numerical study, most of them positivity and maximality criteria,
and taking into account those earlier 271 cases where elimination was very
easy, a typical best detection rate reads, $(98+271)/(160+271) \simeq 85\%$.
\subsection{1-d test functions with four parameters; more statistics; efficiency of the matrix method (Subsec.II B)}

We now consider normalized combinations of the first five even eigenstates
of the harmonic oscillator. Components $c_0$,\,$c_2$,...,\,$c_8$ are random and we
have a set of 4388 cases with both $\psi$ and $\varphi$ positive and, as a
test set for our criteria, 21988 cases where $\varphi$ is partly negative
while $\psi$ remains positive.

For our statistics, we consider in the following ten ``associated
functions'', namely $\psi$, its sign weighted derivatives, 
$(-)^q d^{2q}/dr^{2q}\psi,\, q=1,2,3,4$, then, with $b=1,$ its convoluted form
(9)
 and
its sign weighted derivatives, $(-)^q d^{2q}/dr^{2q} \psi_b,\, q=1,2,3,4$, to
be simultaneously tested. Recall that any of these associated functions is just
a candidate for a positive Fourier transform related to the same $\varphi$.

{\it Bochner test with the 3-matrix} :
A first test consists in the positiveness of the spectrum of the three 
equidistant point Toeplitz matrix \eqref{3matrix}. Typically, we let $r$ run with steps $.025$ until $r=3.5$; this
is sufficient in the present case. As soon as any of the found eigenvalues for
any of the associated functions is negative, the non positivity of $\varphi$
is exposed. This detects 19653 among the 21988 cases of interest, hence an
efficiency of $\simeq 89\% $. If one looks at the remaining 2335 ``rebel''
cases  to see whether the the bound, 
$\psi(r) \ge \psi(0) \cosh(\langle s \rangle r)$, holds, only 14 more cases
show their violation.

{\it Weaker efficiency of the ``cosh test''} :
A second test consists in omitting the ``Bochner test'' and rather starting
by seeing whether the same bound, 
$\psi(r) \ge \psi(0) \cosh(\langle s \rangle r)$, is violated by any among
the ten associated functions. Out of 21988 cases, we find 15887 violations,
hence an efficiency of $\simeq 72\% $ for this mode of negativity detection. 
This leaves 6101 ``rebel'' cases, to be filtered, in turn, by the Bochner test.
We find 3780 additional eliminations. As expected, 19653+14=15887+3780=19667,
the latter number being the number of detections when both test modes are used.

{\it Sensitivity to the convolution parameter} :
We also considered different values of the parameter b, ranging from 
$b=1/5$ to $b=5$. Detections rates are sensitive to $b$. Best rates correspond
to $b$ values between $\simeq 1/2$ and $\simeq 1$, and the choice, $b=1$, 
guided by the exponential decrease in our harmonic oscillator basis, is quite
acceptable.

{\it Weak efficiency of the ``cosh $\pm$ cos'' test } :
For the sake of completeness, the remaining, undetected 2321 cases when $b=1$
were submitted to the test described by Eqs. (14-15). No new detection was 
brought by such bounds. 

{\it Further Bochner tests} :
Then, for the same 2321 cases, the function $\psi$  alone, without its nine
associates, was submitted to the Bochner test, by an ``equidistance'' matrix
again, of order 5 now. A gain of 20 more detections was obtained.

Returning to the initial basis of 21988 cases, this test of $\psi$ alone by
the 5th order Bochner matrix yields  18003 detections, to be compared with
the 19653 results obtained with the third order matrix applied to all ten
associates. A simultaneous test of $\psi$ and $\psi_2$ reaches 18687 detections
and a simultaneous test of $\psi$ and $\psi_b$, with again $b=1,$ returns
19247 detections. This seems to show that a Bochner test on the convoluted 
$\psi_b$ is more useful than a test on the derivative $\psi_2,$ $\psi_2,$... provided 
that $b$ is well chosen. The same fifth order Bochner test applied to $\psi$,
$\psi_2$, $\psi_b$ and $\psi_{b2}$ yields 19941 detections. When all ten
associates are involved, the result reaches 20468, {\it i.e.} $\simeq 93\%$.
Combining this 5th order Bochner test for the ten associates with a 
convexity bound test for the same hardly gives $\simeq 10$ more detections.

{\it Preliminary conclusions on the efficiency of criteria} :
At this stage with this basis of test functions three conclusions stand, namely
i) Bochner tests are more efficient than the cosh tests and similar ones
obtained from the Jensen inequality, ii) ``associate functions'', by
derivations and convolutions, do improve results and iii) good detection
rates, of order 90\% at least, are rather easily available.

\section{Two-dimensional generalization}
\la{id}

We are interested in the same problematics described in the introduction,
applied to the 2-dimensional Fourier case, in the form of the Fourier-Bessel
transform. Indeed, one consider here rotationally invariant partners $\psi(x)$
and $\varphi(k)$,
where $x$ and $k$ are the lengths of two-dimensional vectors $\vec x$ and
$\vec k$, respectively. Typically \cite{kovchegov}, in particle physics,
$\varphi(k)$ corresponds to a gluon distribution in transverse momentum
space $k\equiv \vert\vec k\vert$ and $\psi(x)$ is the quark-antiquark
dipole distribution in transverse size $x\equiv \vert\vec x\vert$,
both distributions being physically required to be positive real functions.

The Bessel-Fourier transforms analogous to formulas (\ref{phi},\ref{psi})
read, 
\eq
\varphi(k)\ &=&\ \int_{0}^{+\infty}\!\! xdx\ J_0(kx)\  \psi(x)\, ,
\la{phiJ}\\
\psi(x)\ &=&\ \int_{0}^{+\infty}\!\! kdk\ J_0(kx)\  \varphi(k) 
\la{psiJ}\, 
\eqx
Conditions i)-v) extend without difficulty, {\it mutatis mutandis}, to this
two-dimensional, and
more precisely radial, situation. The following Subsections are therefore
restricted to only those considerations which reflect the change of
dimensionality.

\subsection{Matrix method}

The Bochner theorem is known to apply to higher dimensional Fourier transforms.
Hence, applying it on the abscissa axis $\vec x = (x,0)$ for the rotational invariant function $\psi(x)$, one is led to consider the matrix method applied to the following set of Toeplitz matrices 
\ba \begin{pmatrix}
\psi(0)& \psi(x)& \psi(2x) & ... & \psi[(n\!-\!1)x]\\
\psi(x)& \psi(0)& ...& ...& \psi[(n\!-\!2)x]\\
...    &     ...& ...&          ...& ...\\
 \psi[(n\!-\!1)x]& \psi[(n\!-\!2)x]& ...& \psi(x)& \psi(0)\\ 
\end{pmatrix}\ .
\la{Nd2matrix}
\ea
Looking for positive definiteness of matrices \eqref{Nd2matrix}, or its
failure, then follows the same path as in the one-dimensional case.

\subsection{Analyticity method}

{\it {\bf Basic lower bound}}
\bigskip

Following the same method as for the one-dimensional Fourier transform, one
defines  the probabilistic formulation of a positive Fourier-Bessel transform
as
\ba
kdk\ {\cal P}(k) \equiv \, 
kdk\ {\va(k)}/\psi(0)\quad \Rightarrow\quad 
\psi(x)= 
\ \psi(0)\ \cor{J_0(kx)} \, ,
\la{shortJ}
\ea 
where $\psi(0)=\int_0^\infty kdk\ \va(k)$ provides the normalization.

For the same reasons as for the 1-d case, the analytic continuation 
$x\!\to\!ix$ leads to the following Jensen inequality,
\ba
\psi(ix)\ = \psi(0)\ \cor{\!I_0(kx)\!}\ \ge \ \psi(0)\ I_0(x \cor{\!k\!})\ =
\ \psi(0)\ I_0 \left[x\!  \int_0^\infty\!\! k{dk}\  k\va(k)/\psi(0) \right],\ 
\quad \forall x \in [0,\infty[\ ,
\la{inequalityJ}
\ea
where the convex function $I_0(u) \equiv J_0(iu)$ is the  Bessel function of
the second kind.

As in the previous section,  the next step is to make from 
Eq. \eqref{inequalityJ} a self-content equation in terms of $\psi$. For this
sake, one introduces the  Laplace transform,
\ba
\chi(x)=  \int_0^\infty \!\! kdk\ e^{-kx}\ \va(k)\ = \ \int_0^\infty \!\! kdk
\ e^{-kx}\int_0^\infty \!\! x'dx'\  \psi(x')\, J_0(kx')\ .
\la{laplacebisJ}
\ea
Exchanging the integrals, while keeping $x > 0$, strictly positive in order
to preserve the convergence of the integration first over $k$ which can be
done analytically, one finds,
\ba
\chi(x)\ =   \ \int_0^\infty \!\! x'dx'\  \psi(x') \int_0^\infty\!\! kdk\  
e^{-kx}\  J_0(kx') = x \int_0^\infty \!\!\f {x'dx'}{(x^2+x'^2)^{3/2}}
\ \psi(x')\ .
\la{laplacebisbisJ}
\ea
An integration by part, similar to that used in subsection II C,
returns the derivative of $\chi$ for all strictly positive  $x$,
\ba
\f {d\chi}{dx}(x)\ =\  \int_0^\infty \!\!x'dx'\ 
\f{2x^2-x'^2}{(x^2+x'^2)^{5/2}} \ \psi(x')\ =\ -\int_0^\infty \!\! dx'\ 
\f{d\psi}{dx'}(x')\ \f{x'^2}{(x^2+x'^2)^{3/2}}\ ,
\la{laplaceJ}
\ea
where the last expression, obtained by another integration by part, is now
regular at $x=0$ and thus  a faithful representation of the function 
$d\chi/dx$ for all $x \ge 0$. One then gets for the required moment,
\ba
\int_0^\infty\!\!\!\!
 k{dk}\  k\va(k) = - \f {d\chi}{dx}(x\!=\!0)\ = \ - \int_0^\infty \!\!
\f {dx'}{x'}\ \f {d\psi}{dx'}(x')\ = \  \int_0^\infty \!\!\f {dx'}{x'^2}\ 
\left(\psi(0)-\psi(x')\right)\ .
\la{laplace4J}
\ea
Interestingly enough,  this is  the same expressions, except for a 
coefficient, as (\ref{finalpart}) obtained in the 1-d case.

Inserting the result \eqref{laplace4J} into the inequality 
\eqref{inequalityJ} yields,
\ba
\psi({ix})\ \ge\ \psi(0)\ I_0\left[ {x}\!  \int_0^\infty\!\,
\!\f {dx'}{x'^2}\, \left(1-\f{\psi(x')}{\psi(0)}\right)\right], \quad  
\forall { x} \in [0,\infty]\ .
\la{constraintJ}
\ea

{\it \bf Hierarchy of further bounds}
\bigskip


Let us introduce the hierarchy of positive functions,
\ba
\va_{p}(k)\ \equiv\ k^{p}\va(k)\ ; \quad \va_{0}(k)\equiv \va{(k)}\, .
\la{hieraJ}
\ea
Following the same approach as in section \ref{Introd}, we introduce the
probability distributions,
\ba
kdk\ {\cal P}_p(k) \equiv kdk\ \f {\va_p(k)}{\int_0^\infty \, kdk\ \va_p(k)}\ .
\la{probahieraJ}
\ea
We also introduce the 2-dimensional radial Laplacian operator $\Oo$ acting on 
$\psi(x)$. It leaves the Bessel functions $J_0(kx)$, $I_0(kx)$ invariant up
to factors, namely,
\ba
\Oo \left[\psi(x)\right]\ \equiv\ - \f 1x\f d{dx}\left(x\f d{dx}\right)
\left[\psi(x)\right]\quad \Rightarrow\quad\Oo\left[J_0(kx)\right]\ =
\ k^2\ J_0(kx) \, .
\la{operatorJ}
\ea
This operator $\Oo$ plays in our 2-dimensional problem the role of the
ordinary double derivative in 1-d. Then, for even $p=2q$, we now define a new
function $\psi_{2q}(x)$ from the multiple $\Oo$-derivatives of $\psi(x)$,
\ba
\psi_{2q}(x)\ \equiv \ \left[\Oo\right]^q \psi(x)\ =\ \int_{0}^{+\infty} 
kdk\ J_0(k x)\  k^{2q}\va(k)\ = \ \int_{0}^{+\infty} kdk\  J_0(kx)\ 
\va_{2q}(k)\, ,
\la{shorthieraJ}
\ea
by acting  on  the Bessel-Fourier transform formula \eqref{psiJ}. 
In particular, we get for an even moment of $\varphi$, of order $2q$, namely
the zeroth order moment of $\va_{2q}$,
\ba
\psi_{2q}(0)\ =  \ \left.\left[\Oo\right]^{q}\psi(x)\right\vert_{x=0}\ =\  
\int_{0}^{+\infty}\!\! kdk\ \va_{2q}(k)\, .
\la{evenmomentsJ}
\ea
Now, extending the Jensen inequality \eqref{inequality} to the probability 
distributions \eqref{probahieraJ}, we get for the positive functions 
$\psi_{2q}(ix)$ the following hierarchy of inequalities,
\ba
\psi_{2q}(ix)\ = \ \psi_{2q}(0) \ \cor{\!I_0(xk)\!}_{\Po_{2q}} \ \ge\ 
\psi_{2q}(0)\ I_0\left\{\cor{\!xk\!}_{\Po_{2q}}\right\}\ = \ \psi_{2q}(0)\ 
I_0 \left\{x\ \f { \int_0^\infty kdk\ \va_{2q+1}(k)}{ \int_0^\infty\!\!
k{dk}\ \va_{2q}(k)} \right\} \ ,
\la{inequalityhieraJ}
\ea
where,
\ba
\psi_{2q}({ix}) =  \int_{0}^{+\infty}\!\! kdk\  I_0(xk)\  \va_{2q}(k)\, ,
\la{constrainthieraJ}
\ea
is the analytic continuation of $\psi(x)$ onto the imaginary axis.

Since $\psi_{2q}$ is just another function assumed to have a positive
Fourier transform, we can at once take advantage of Eq. \eqref{constraintJ}
and obtain,
\ba
\psi_{2q}(ix)\ \ge\ \psi_{2q}(0)\ I_0 \left\{x\, \int_0^\infty \, 
\f{dx'}{x'^2}  \left(1-\f{\psi_{2q}(x')}{\psi_{2q}(0)}\right)\right\} \ .
\la{inequalityhieraJJ}
\ea
\eject
{\it \bf Convolution}
\bigskip

For any width $b$ the product, $\exp[-k^2/(2b^2)]\ \varphi(k)$, retains the
assumed positivity of $\varphi$ and converts the partner $\psi$ into the
result of a convolution,
\begin{equation}
\psi_b(x)=b^2/(2 \pi) \int_0^{2\pi} d\theta \int_0^{\infty} x'dx'\,
\exp[-(x^2+x'\,^2-2 x x' \cos \theta)\, b^2/2]\ \psi(x')\, ,
\end{equation}
an obvious generalization of Eq. \eqref{convolu}. The angle $\theta$ is that
between $\vec r$ and $\vec r'$ and the angular integration yields,
\begin{equation}
\psi_b(x)=b^2\, \exp[-b^2 x^2/2] \int_0^{\infty} x'dx'\, 
I_0(b^2 x x')\, 
\exp[-b^2 x'\, ^2/2]\ \psi(x')\, .
\end{equation}

\bigskip

Additional inequalities are easily obtained because the Taylor series of 
$J_0$
has the same alternating sign structure as that of $cos$. The combinations,
$J_0(x)+J_0(ix)$ and $J_0(x)-J_0(ix)$, therefore, are convex and concave, 
respectively. Generalizations of inequalities (14)-(19) to the present radial,
2-d situation are obvious. An extension of relations (14)-(19) to this
hierarchy of functions $\psi_{2q}$, and to the corresponding functions
associated with $\psi_b$, is trivial.
\bigskip

\subsection{Numerical investigations for the Bessel transform}
We generated 10127 test functions $\psi(x)$ by randomly mixing the following
5 functions,
$\sqrt{2}\, e^{-x^2/2}$, $\sqrt{2}\, e^{-x^2/2} (1 - x^2)$, 
 $e^{-x^2/2} (2 - 4 x^2 + x^4)/\sqrt{2}$, 
 $e^{-x^2/2} (6 - 18 x^2 + 9 x^4 - x^6)/(3 \sqrt{2})$, 
 $e^{-x^2/2} (24 - 96 x^2 + 72 x^4 - 16 x^6 + x^8)/(12 \sqrt{2})$. This is an
 orthonormal basis with the measure $xdx$, $0\le x < +\infty,$ using 
Laguerre polynomials. The basis states are 
eigenstates of the Fourier-Bessel transform, with eigenvalues 
$\{1,-1,1,-1,1\}$, respectively. Then our random mixtures were screened
 so that each $\psi$ be square normalized, positive, with its maximum at the
origin, while $\varphi$ has some negativity, to be detected.

We implemented the ``matrix test'' at orders 5, 8, 9 and 10 for $\psi$ alone, 
for simplicity,
neglecting its associated derivative and convoluted functions. Detection 
rates were $\simeq 29\%$, $39\%$, 
$41\%$, $43\%$, respectively, compared to 
$\simeq 69\%$, $82\%$, $84\%$, $86\% ,$ for the 1-dimensional case again for $\psi$ alone. 
These 2-d detection rates are less satisfactory than the detection
rates obtained with our 1-d test set, but still acceptable, because of our
screening for cases with a maximum at the origin only.

The analyticity test for $\psi$ alone returned hardly a $\simeq 3\%$ 
detection rate, which is weak and similar  to 
$\simeq 5\%$ for the 1-dimensional case.
 This is confirmed by combining the matrix
test at order 10 with this analyticity test. Except for one case, those $14\%$ 
functions undetected by the matrix test also escape the analyticity test.

\section{Conclusion and outlook}
The constructive characterization of real functions with positive Fourier 
transforms in one and two-dimensional (radial) dimensions has been analyzed here 
with 
two different methods. One, the ``analyticity method'', makes use of  analytic 
continuation combined with convexity 
properties (Jensen inequalities).  The other,
 the ``matrix method'', deals with finite  Toeplitz matrices testing the Bochner
 theorem through tests of positive definiteness. 

We test  these methods on a set of random mixtures of functions combining a 
Gaussian with
 polynomials  
forming a finite, but  of sizable dimension, orthogonal basis. We estimate and
 compare the efficiencies of
 both methods. 

In one dimension, the matrix method reaches 
a $86\%$ efficiency for the test on the function $\psi$ alone, and reaches easily
 more than $93\%$ for the combined test including derivatives. This is to be
 compared with  $43\%$ for the analyticity method and  $85\%$ at most, 
after
 combining with suitable convolutions and differentiation preserving the 
positivity of the Fourier
 transform. All in all the matrix method seems at present quite successful and 
more directly efficient than the 
analyticity one.

In the two-dimensional (radial) case, the situation is less satisfactory for the 
matrix method. It  reaches a $43\%$ efficiency under the same conditions, but is 
still probably not too difficult to improve quantitatively. The analyticity 
method, by contrast, is at the present stage much ineffective, with only $\simeq 
3\%$ detection for the same set of non positive Fourier-Bessel transforms. This 
method  requires a qualitative improvement to be competitive; a good subject for 
future research. 

By way of an outlook,  there are possible directions for improvement.

Although our sets of test functions are reasonably big ($10^4$ typically) and
 diversified, we feel that our conclusions should be strengthened by the
 consideration of a larger basis and  not necessarily weighted by Gaussians, with more allowed 
configuration mixing. However, we may already define quite a few new research 
directions on the same line, namely:

- In two dimensions, it is clear that a vectorial version of the matrix model is 
to be investigated, by comparison with the one-dimensional projection used in 
 section IV.

- The analyticity method,  despite its mathematical simplicity based on generalized convexity properties, seems at its
 present  stage suffering from a  deficit of efficiency. The question remains of 
improved convexity tests.

- Considering the matrix method and its growing accuracy with matrix dimension, 
it is tempting to consider the possibility of using Toeplitz matrices of large 
rank.

- At last but not least, the full constructive characterization of positive 
positive-definite functions is yet to be reached, hoping that the steps made in
 the present paper may lead to this goal.

\section*{Acknowledgements}
We want to thank Philippe Jaming (Mathematical institute, Bordeaux I U.) for
 some fruitful insights on positive positive-definite 
functions, and Bertrand Eynard for interesting mathematical remarks on the 
subject.


\begin{thebibliography}{0}

\bibitem{gipe} 
  B. G. Giraud and R. B. Peschanski,
  ``On positive functions with positive Fourier transforms,''  Acta Phys.
\ Polon.\ B {\bf 37}, 331 (2006)  [math-ph/0504015].

\bibitem{levy}
P.\ L\'evy: Fonctions caract\'eristiques positives [Positive 
characteristic functions]. {\it Comptes Rendus Hebdomadaires
des S\'eances de l'Acad\'emie des Sciences, S\'erie A,
Sciences Math\'ematiques} {\bf 265} (1967) 249-252. [in French]
Reprinted in: {\it \OE uvres de Paul L\'evy. Volume III.
El\'ements Al\'eatoires}, edited by
D.\ Dugu\'e in collaboration with  P.\ Deheuvels, M.\ Ib\'ero.
Gauthier-Villars \'Editeur, Paris, 1976, pp.\ 607-610.

\bibitem{kovchegov}
Y.~V. Kovchegov,
\newblock Phys. Rev. {\bf D60}, 034008 (1999); Phys. Rev. {\bf D61}, 074018
(2000).

\bi{math}
P. Jaming, M. Matolcsi and S.G. R\'ev\'esz,
``On the extremal rays of the cone of positive, positive definite
functions,'' J. Fourier Anal. Appl.,  1, 561 (2009).
\newline
A. Hinrichs, J. Vybíral,
``On positive positive-definite functions and Bochner's Theorem,''
 J. Complexity 27 , 264, (2011).
 \newline
 D. Heck, T. Schl\"omer and O. Deussen,
``Blue noise sampling with controlled aliasing.''
ACM Trans. Graph. 32,3,25:1, (2013).

\bi{phys}
S. Pigolotti, C L\'opez, E Hern\'andez-Garc\' \i a,
``Species clustering in competitive Lotka-Volterra models,''
Phys. Rev. Lett.,  98, 258101 (2007). 2007
\newline
F. Baldovin, A.L. Stella,
``Central limit theorem for anomalous scaling due to correlations,''
Phys. Rev. E, 75(2), 020101, (2007). 
\newline
T. Lappi  
``Small x physics and RHIC data,''
Int. J. Mod. Phys. E 20, 1, (2011) .
\newline
B. Deissler, E. Lucioni, M. Modugno, G. Roati, L. Tanzi, M. Zaccanti,
M. Inguscio, and G. Modugno,
 "Correlation function of weakly interacting bosons in a disordered
lattice," 
New J. Phys. 13, 023020 (2011).
\newline
C. Cotar, G Friesecke and B. Pass,
``Infinite-body optimal transport with Coulomb cost,''
arXiv:1307.6540, 2013. 
\newline
  B.~G.~Giraud and S.~Karataglidis,
  ``Algebraic Density Functionals,'' 
 Phys.\ Lett.\ B {\bf 703}, 88 (2011).

\bi{eynard}
B. Eynard, private communication.

\bibitem{general}
For a comprehensive list of references, see  K.~Scharnhorst,
``A Grassmann integral equation,''
J.\ Math.\ Phys.\  {\bf 44}, 5415 (2003)
[arXiv:math-ph/0206006], in particular the references [111] to [125].

\bibitem{gelfand}
See, for instance, I.M. Gel'fand and N.Ya. Vilenkin (1968), {\it Generalized Functions}, Vol. IV,
Academic Press, New-York and London.

\bibitem{lafforgue} We thank T. Lafforgue for a fuitful private communication on convexity properties.

\bi{Toeplitz}
Many textbooks on Toeplitz matrices are available. A convenient introduction is \\
R.M. Gray,
``Toeplitz and Circulant Matrices: A Review,'' in the book
``Foundations and Trends in
Communications and Information Theory''
Vol. 2, No 3 (2006) 155–239, (Now, editor),\\
http://lthiwww.epfl.ch/~leveque/Matrix/gray.pdf .



\bi{jensen}
J. L. W. V. Jensen,  ``Sur les fonctions convexes et les in\'egalit\'es entre
les valeurs moyennes'', Acta Mathematica 30 (1)(1906) 175.

\end{thebibliography}
\end{document}